# Anisotropic thermal conductivity tensor measurements using beam-offset frequency domain thermoreflectance (BO-FDTR) for materials lacking in-plane symmetry

Lei Tang[a] and Chris Dames[a,b,*]

[a]Mechanical Engineering, University of California, Berkeley, Berkeley, California 94720, USA

[b]Materials Sciences Division, Lawrence Berkeley National Laboratory, Berkeley, California 94720, USA

*cdames@berkeley.edu

## Abstract

Many materials have anisotropic thermal conductivity, with diverse applications such as transistors, thermoelectrics, and laser gain media. Yet measuring the thermal conductivity tensor of such materials remains a challenge, particularly for materials lacking in-plane symmetry (i.e., transversely anisotropic materials). This paper demonstrates thermal conductivity tensor measurements for transversely anisotropic materials, by extending beam-offset frequency-domain thermoreflectance (BO-FDTR) methods which had previously been limited to transversely isotropic materials. Extensive sensitivity analysis is used to determine an appropriate range of heating frequencies and beam offsets to extract various tensor elements. The new technique is demonstrated on a model transversely anisotropic material, x-cut quartz (<110> $\alpha$-SiO$_2$), by combining beam offset measurements from different sample orientations to reconstruct the full in-plane thermal conductivity tensor. The technique is also validated by measurements on two transversely isotropic materials, sapphire and highly oriented pyrolytic graphite (HOPG). The anisotropic measurements demonstrated very good self-consistency in correctly identifying isotropic directions when present, with residual anisotropy errors below 4% for sapphire and 2% for HOPG and quartz. Finally, a computational case study (simulated experiment) shows how the arbitrary in-plane thermal conductivity tensor of a fictitious material with high in-plane anisotropy can in principle be obtained from only a single sample orientation, rather than multiple orientations like the experiments on x-cut quartz.

# 1. Introduction

The thermal conductivity of many anisotropic materials remains to be investigated. In addition, some nominally isotropic materials with oriented defects like dislocations [1] can exhibit highly anisotropic thermal conductivity. Furthermore, such anisotropic materials can find many applications including as transistors [2], thermoelectrics [3], [4], high temperature superconductors [5], and laser gain media [6], [7]. Therefore, the capability of anisotropic thermal conductivity measurements is of significant fundamental and practical importance. Yet, these measurements remain challenging, particularly for materials lacking in-plane symmetry.

So far, the most common techniques used to measure anisotropic thermal conductivities are capitalizing on the electro-thermal and optical pump-probe methods, including the 3-omega method [8], [9], [10], microfabricated suspended devices [11], [12], frequency-domain thermoreflectance (FDTR) [13], [14], [15], [16], [17], and time-domain thermoreflectance (TDTR) [18], [19], [20], [21], [22]. Compared to FDTR and TDTR, the 3-omega and microfabricated device methods are not scannable for spatial property mapping and require more complicated microfabrication, which limit their applications. As compared to TDTR, FDTR is more cost-effective and simpler to implement as it doesn't require an ultrafast pulsed laser or a delay stage. Due to these advantages, the development of FDTR is important for anisotropic thermal conductivity measurements.

Conventional FDTR with co-aligned continuous wave (CW) laser beams has been demonstrated for measuring the in-plane and cross-plane thermal conductivities of bulk materials [13], thin films [14], and multilayers [15]. Recently, Rodin and Yee [16], Rahman et al. [17], and Qian et al. [23]

used offset pump and probe beams, known as beam-offset FDTR (BO-FDTR), to increase the sensitivity and thus accuracy of the in-plane thermal conductivity measurements. Yet, all those measurements have been limited to transversely isotropic materials, meaning the thermal conductivity is the same in all directions parallel to the sample surface.

Here, we extend the previous BO-FDTR capabilities [16], [17], [23] to measure the thermal conductivity tensor for the more general case of materials lacking in-plane symmetry. The mathematical framework for BO-FDTR of materials with arbitrary anisotropy is developed here by adapting related work from the TDTR literature [20], [21]. Extensive sensitivity analysis is used to identify suitable regimes of heating frequency and pump-probe beam offset to extract the various tensor elements. The new technique is first validated by control measurements of two transversely isotropic materials: sapphire, which is weakly anisotropic (cross-plane vs. in-plane), and highly oriented pyrolytic graphite (HOPG), which is highly anisotropic. Then the technique is demonstrated on a transversely anisotropic material, x-cut quartz (<110> $\alpha$-$SiO_2$), by combining measurements in different directions obtained from various sample orientations to reconstruct the thermal conductivity tensor. All obtained results show good agreement with literature values for the thermal conductivity tensors. Further, a purely computational case study is conducted showing how these methods can in principle be extended to measure the in-plane thermal conductivity tensor of a material with a high in-plane anisotropy ratio by performing a beam-offset experiment at a single sample orientation, rather than multiple measurements at various sample orientations as done for the x-cut quartz.

The paper is organized as follows. First, the details of the experimental setup and mathematical model will be discussed. Then thermal conductivity measurements of different samples, namely sapphire, HOPG, and x-cut quartz, will be presented with detailed analysis. Finally, the computational case study will be discussed.

## 2. Experimental Setup

The working principle of FDTR is well established [18], [24], [25]. Briefly, FDTR utilizes an optical pump-probe method to measure the thermal phase lag between the transducer's surface temperature and the modulated pump laser heating. The surface temperature is obtained from the temperature-dependent reflectance (i.e., thermoreflectance) of a CW probe laser. The measured phase lag as a function of heating frequency is then fit to a theoretical model to determine the unknown thermal parameters of interest.

Our FDTR system was acquired from Fourier Scientific LLC and is based on the experimental approach of A. J. Schmidt. The experimental setup is shown in Fig. 1(a) and closely follows Refs. [13] and [14]. The pump and probe lasers (CrystaLaser, CL532-050-LO and DL488-180-O) have wavelengths of 488 nm and 532 nm, respectively. Both lasers have built-in optical isolators to prevent back reflections into the lasers. The pump laser is modulated electrically at frequency $f$ via a lock-in amplifier (Zurich Instruments, HF2LI) and is directed to the sample surface by a dichroic mirror. The samples are coated with a thin gold transducer layer of thickness $t \sim 75$ nm to absorb pump heat more efficiently as well as provide a larger coefficient of thermoreflectance at the probe wavelength ($\sim 2 \times 10^{-4}$ $K^{-1}$ [26]). The probe laser passes through a polarizing beam splitter (PBS) which splits it into two beams. One probe beam reaches the sample surface close to the pump spot location and then is reflected to the photodiode P1 in a balanced photodetector. The other probe beam is sent directly to the photodiode P2. This scheme facilitates subtraction of common-mode noise by taking the difference between the signals from P1 and P2. The half waveplate just before the PBS is adjusted to achieve fine balancing between P1 and P2. Also, two bandpass filters are placed in front of the detector to block other irrelevant signals such as scattered pump light which could be detrimental to the measurements. Then the thermal phase lag, $\phi$, defined as the difference between the reflected probe phase and the pump phase, is obtained via the lock-in amplifier. Other miscellaneous non-thermal phase shifts such as caused by the separate optical pump and probe paths and the driving electronics, were nulled out through control experiments on standard samples whose thermal conductivities had already been carefully measured and analyzed by the vendor using their previous well-established FDTR system. All measurements in this study are performed at ambient temperature. Other details of this experimental setup are well explained in Refs. [13] and [14].

One important extension in the current system as compared to the previous setup [13], [14] is the addition of a pair of high-resolution motorized actuators to tilt the dichroic mirror, thereby moving the pump spot precisely. This spot motion was calibrated using a precisely patterned sample, and we found that the uncertainty of the pump spot movement plays a negligible role as compared to the other uncertainty contributions discussed below. Although in the actual experimental apparatus the probe location is fixed and the pump is deflected, for mathematical convenience in the rest of this paper we locate the origin of the xyz coordinate system at the center of the pump, as indicated in Fig. 1b(right). These two configurations are equivalent since it is only the relative distance between pump and probe that matters.

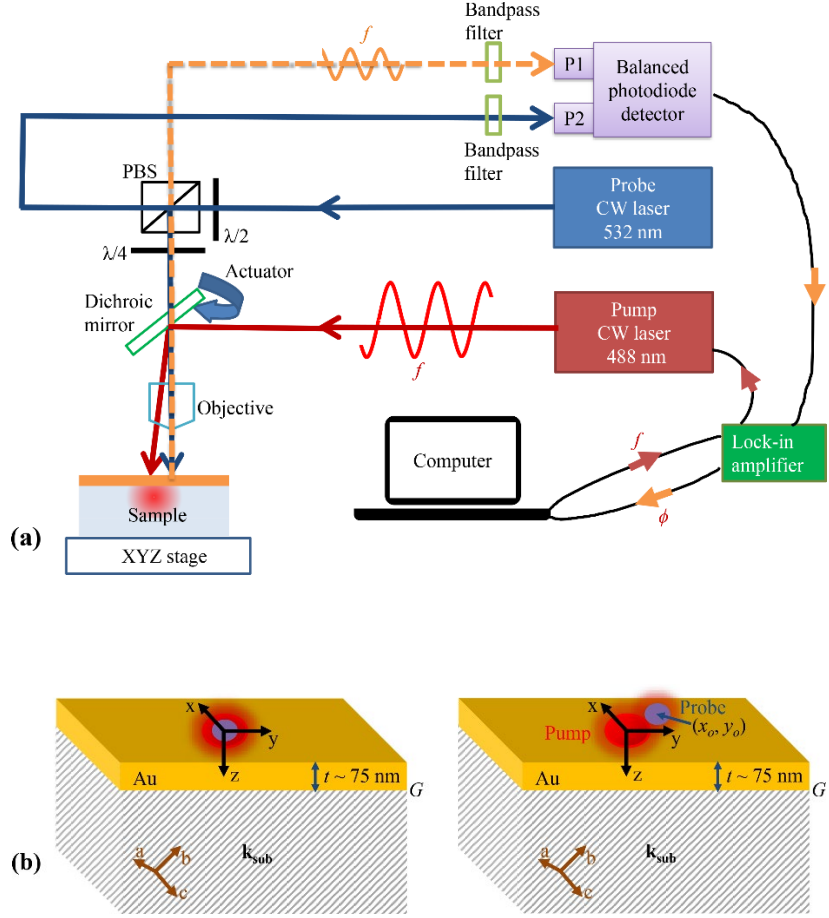

Figure 1. (a) Experimental setup. This is similar to the apparatus described in Refs. [13], [14], with the addition of the actuated dichroic mirror for xy-scanning of the pump spot location. A lock-in amplifier is used to modulate the pump laser and detect the signal from a balanced photodetector consisting of two photodiodes P1 and P2. The photodiode P1 measures the reflectance of the probe beam from the sample surface while the signal from P2 is used to cancel common-mode noise. (b) Sample configurations for FDTR with (left) co-aligned and (right) offset beams. The xyz coordinate system is used in the analytical model and is aligned with its origin centered under the pump laser and the xy plane located at the top of the transducer layer. This is different from the sample principal axes along the a, b, and c directions, which in general may be aligned arbitrarily.

The unknown parameters of interest are the thermal interface conductance between the Au transducer layer and sample surface, *G*, and the thermal conductivity tensor of the sample, **k**$_{sub}$.

## 3. Mathematical Model

3.1 Multilayer heat conduction equation

The heat flux vector **q** is related to the thermal conductivity tensor and temperature gradient by Fourier's law [27],

$$\mathbf{q} = -\mathbf{k}\nabla T, \tag{1}$$

where the most general thermal conductivity tensor **k** in an arbitrary cartesian coordinate system is expressed as

$$\mathbf{k} = \begin{bmatrix} k_{xx} & k_{xy} & k_{xz} \\ k_{yx} & k_{yy} & k_{yz} \\ k_{zx} & k_{zy} & k_{zz} \end{bmatrix}. \tag{2}$$

In the absence of magnetic fields the thermal conductivity tensor is symmetric ($k_{ij} = k_{ji}$) [28] and therefore at most 6 independent variables are required to describe the most general anisotropic **k** in this work. We also require the sample to be homogeneous (i.e., **k** is independent of position). For the case of nonhomogeneous materials, scanning the sample can potentially be used to obtain thermal conductivity mapping [13], which is outside the scope of this study.

The heat diffusion equation for an anisotropic medium without heat generation is

$$k_{xx}\frac{\partial^2 T}{\partial x^2} + k_{yy}\frac{\partial^2 T}{\partial y^2} + k_{zz}\frac{\partial^2 T}{\partial z^2} + 2k_{xy}\frac{\partial^2 T}{\partial x \partial y} + 2k_{xz}\frac{\partial^2 T}{\partial x \partial z} + 2k_{yz}\frac{\partial^2 T}{\partial y \partial z} = C\frac{\partial T}{\partial t}, \tag{3}$$

where *C* is the volumetric heat capacity of the medium. As a special case, if the x-, y-, and z-axes of the coordinate system are respectively aligned along the principal a, b, and c-axis directions of the material crystal with their respective principal thermal conductivities $k_a$, $k_b$, and $k_c$, all off-diagonal terms of **k** are exactly zero [27], such that the left-hand side of Eq. (3) simplifies to only three terms with $k_{xx} = k_a$, $k_{yy} = k_b$, and $k_{zz} = k_c$.

For time-periodic forcing at frequency $f$, the anisotropic heat equation can be solved in the frequency domain using spatial Fourier transforms, and the solutions are well established in the TDTR literature [20], [21]. For gaussian pump and probe beams offset by a distance $(x_o, y_o)$ as indicated in Fig. 1(b), the relevant probe-weighted-average surface temperature is

$$\Delta T = \int_{-\infty}^{\infty}\int_{-\infty}^{\infty} H(f,\eta,\xi)\exp[-\frac{\pi^2}{2}(w_0^2 + w_1^2)(\eta^2 + \xi^2)]\exp[i2\pi(\eta x_o + \xi y_o)]d\eta d\xi, \quad (4)$$

where the temperature response function $H$ accounts for all thermal properties including the thermal conductivity tensor elements and volumetric heat capacities of both substrate and transducer layer, as well as the thermal interface conductance $G$. The full expression for $H$ is given in [21]. The symbols $w_0$ and $w_1$ denote the pump and probe $1/e^2$ radii. Finally, the thermal phase lag is calculated as

$$\phi = \tan^{-1}\frac{\text{Im}(\Delta T)}{\text{Re}(\Delta T)}. \quad (5)$$

In general the strategy is to measure $\phi(f)$ and fit it with the modeled $\phi(f)$ by varying a small number of unknown parameters, such as one or a few elements of the sample's thermal conductivity tensor.

3.2 Sensitivity analysis

A sensitivity analysis can be used to visualize how sensitive the measured phase lag is with respect to a change in a specific parameter. The sensitivity is defined as

$$S_\gamma = \frac{\partial \ln(\phi)}{\partial \ln(\gamma)} \tag{6}$$

where $\gamma$ denotes some parameter of interest. For unknowns which we desire to determine by fitting, such as elements of the **k** tensor, large sensitivity is desired to minimize the uncertainty in the fit parameter. In addition, it is difficult to independently determine two fit parameters if both have similar trends in their $S_\gamma(f)$ functions; formally the effects of two unknown parameters $a$ and $b$ are indistinguishable in a measurement if $\frac{S_a(f)}{S_b(f)} = constant$ throughout the range of the experiment. Therefore, it is important to conduct a sensitivity analysis to select the appropriate experimental regimes such as heating frequency range and pump-probe beam offset(s). The detailed sensitivity analysis for each sample will be discussed in Section 5.

3.3 Uncertainty analysis

The uncertainties arising from all the input parameters can be calculated as [24], [29]

$$\Delta \gamma = \left[\sum_\beta (\Delta \gamma_\beta)^2\right]^{1/2} = \gamma \frac{\partial \ln \gamma}{\partial \ln \phi} \left[\sum_\beta (S_\beta \frac{\Delta \beta}{\beta})^2\right]^{1/2}, \tag{7}$$

where $\Delta \gamma$ is the total uncertainty in some parameter $\gamma$ of interest, $\Delta \gamma_\beta$ is the contribution due to parameter $\beta$, and $\beta \neq \gamma$ represents a list of all other input parameters used to evaluate Eq. (4), each with their own uncertainties $\Delta \beta$. In this work the principal unknowns are the thermal conductivity tensor of the sample and the thermal interface conductance between the Au transducer layer and the sample. All other input parameters are considered known, with nominal values and uncertainties (68% confidence intervals) listed in Table 1. Note that we neglect the uncertainty in beam offsets ($\Delta x_o$, $\Delta y_o$) as it is sufficiently small as to have no noticeable impact on the measurements of **k** and $G$. To evaluate Eq. (7), we closely follow the numerical calculation method described in Ref. [29], which will be briefly explained in Section 5.1.

Table 1. Nominal values of the input thermal parameters with their ± uncertainties (68% confidence intervals). $t_{Au}$ was measured using AFM, $k_{Au}$ was determined from the measured electrical resistivity using the Wiedemann-Franz law, and $w_0$ and $w_1$ were measured using a knife-edge technique.

| Input Parameter | Nominal value | Uncertainty |
| --- | --- | --- |
| $k_{Au}$ | 260 W/m-K | 2% |
| $t_{Au}$ | 75 nm | 1% |
| $C_{Au}$ | 2.49 MJ/m$^3$-K [30] | 2% |
| $C_{sample}$ | $C_{sapphire}$ = 3.09 MJ/m$^3$-K [31]<br>$C_{HOPG}$ = 1.59 MJ/m$^3$-K [32]<br>$C_{quartz}$ = 1.96 MJ/m$^3$-K [33] | 2% |
| $w_0$ | 4.2 μm | 1.5% |
| $w_1$ | 2.0 μm | 2% |

To understand how each input parameter influences the uncertainty of the measurements of **k** and $G$, the sensitivities of these parameters as functions of frequency, $S_\beta(f)$, are plotted in Fig. 2, using sapphire as an example. Both $S_\beta$ and $\Delta\beta/\beta$ affect the uncertainty of the measurements of **k** and $G$ based on Eq. (7). Yet, since the uncertainties of all the input parameters $\Delta\beta/\beta$ have relatively similar values (within a factor of 2 of each other; see Table 1) and are constant across all measurements in this study, the much more significant information is conveyed through the sensitivities $S_\beta(f)$. As can be seen in Fig. 2, $\phi$ is only sensitive to the heat capacity and thickness of the Au at high frequencies due to their small penetration depths, while the sensitivities to the heat capacity of sapphire and the pump radius have relatively high values for all considered frequencies. These observations hold true both with and without beam offset, although the absolute sensitivity values are generally decreased by introducing the beam offset. A larger uncertainty associated with the measurements of the unknown parameters (like $G$ and **k**) will be observed if the input parameters have higher sensitivity values. Therefore, the uncertainty of the measurements will be dominated by those four parameters ($C_{Au}$, $t_{Au}$, $C_{sapphire}$, and $w_0$) for sapphire. We have repeated analogous sensitivity calculations for HOPG and quartz (details omitted for brevity) and the same conclusions are found about the four dominant parameters.

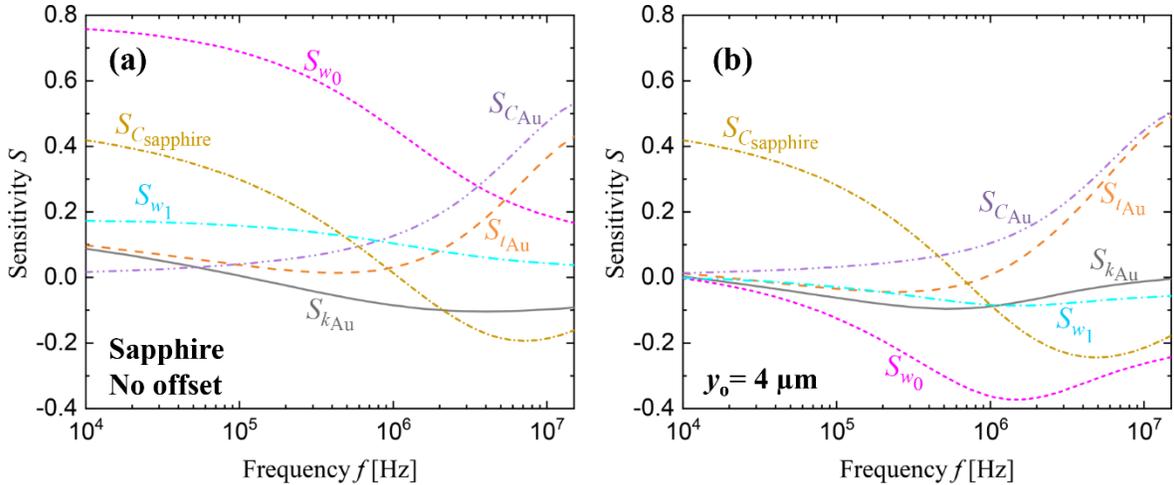

Figure 2. Sensitivities of $\phi(f)$ to the input parameters from Table 1 for sapphire with (a) no beam offset and (b) 4 μm beam offset in the y direction.

## 4. Sample preparation

All samples except HOPG were carefully cleaned by sonicating in acetone, isopropanol (IPA), and deionized water for ~5 minutes each and subsequently dried using a nitrogen blow gun before the transducer layer deposition. HOPG was exfoliated a few times using Scotch tape to smooth the rough surface. Then, a 75 nm Au layer was directly deposited on all samples by electron beam evaporation without any adhesion layer. To ensure the uniformity of the Au layer, the deposition rate typically was kept below 1 Å/s. After the deposition, the thickness of the Au layer was determined by AFM. The thermal conductivity of the Au layer was then obtained using four-point probe measurements of electrical conductivity and applying the Wiedemann-Franz law with the free-electron Lorenz number assuming isotropic transport properties. Note that such AFM and electrical conductivity measurements were only performed on x-cut quartz, and we applied the same transducer property results to sapphire and HOPG samples since the deposition process was completed in the same batch for all samples.

## 5. Results and Discussion

5.1 Validation using transversely isotropic materials: Sapphire and HOPG

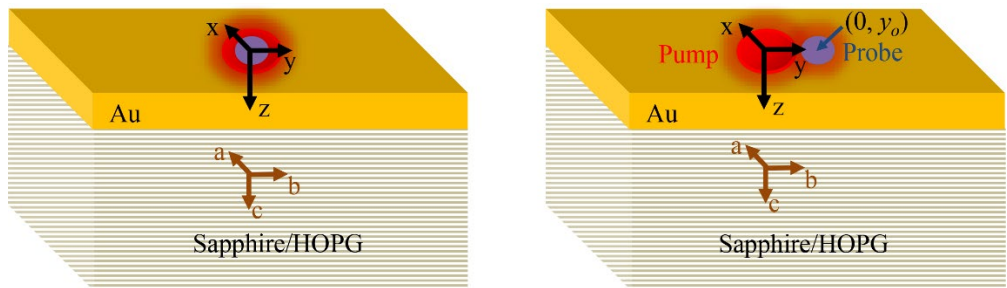

Figure 3. Schematic for sapphire and HOPG samples with co-aligned (left) and offset (right) beams. In both cases the x, y, and z-axes of the optical system are aligned parallel to the principal a-, b-, and c-directions of the sample crystal structure, respectively.

First, the experimental system and analytical model are validated by measurements on two prototypical transversely isotropic materials, sapphire and HOPG. These samples both have hexagonal crystal structures, so that from symmetry it is required that $k_a = k_b$ [34]. Since as shown in Fig. 3 these samples are aligned such that the transducer layer (which also defines the xy plane of the optical coordinate system) is deposited on an ab surface of the sample, these materials are transversely isotropic. Therefore, in this xyz coordinate system the thermal conductivity tensors of these materials have no off-diagonal terms, and the principal in-plane thermal conductivities $k_a$ and $k_b$ and cross-plane thermal conductivity $k_c$ are respectively along the x-, y- and z-axis directions and need to be determined. In other words, $k_{xx} = k_{yy} = k_a = k_b$ and $k_{zz} = k_c$. In addition, the thermal conductivity of sapphire was reported to be almost isotropic (the difference between $k_a$ (which $= k_b$) and $k_c$ is only around 6% [35]) and is often approximated as perfectly isotropic [31], [36].

### 5.1.1 Sapphire

FDTR measurements were first conducted on a sapphire sample with no beam offset, corresponding to the situation of Fig. 3(left). The sensitivities of $\phi(f)$ to the principal thermal conductivities and thermal interface conductance are plotted as functions of heating frequency in Fig. 4a. As can be seen, the in-plane thermal conductivities cannot be obtained individually without beam offset as their sensitivities at every frequency are identical ($S_{k_a}/S_{k_b} = 1$), so we make no attempt to determine them from these no-offset measurements. On the other hand, the sensitivities to the cross-plane thermal conductivity and thermal interface conductance are quite distinct from each other ($S_G/S_{k_c} \neq$ constant), particularly at the intermediate and high frequencies, such that they can be determined by measuring the phase lag for this frequency range under this configuration of no beam offset. Therefore, for the zero-beam-offset case, only the fits for $G$ and $k_c$ are unique.

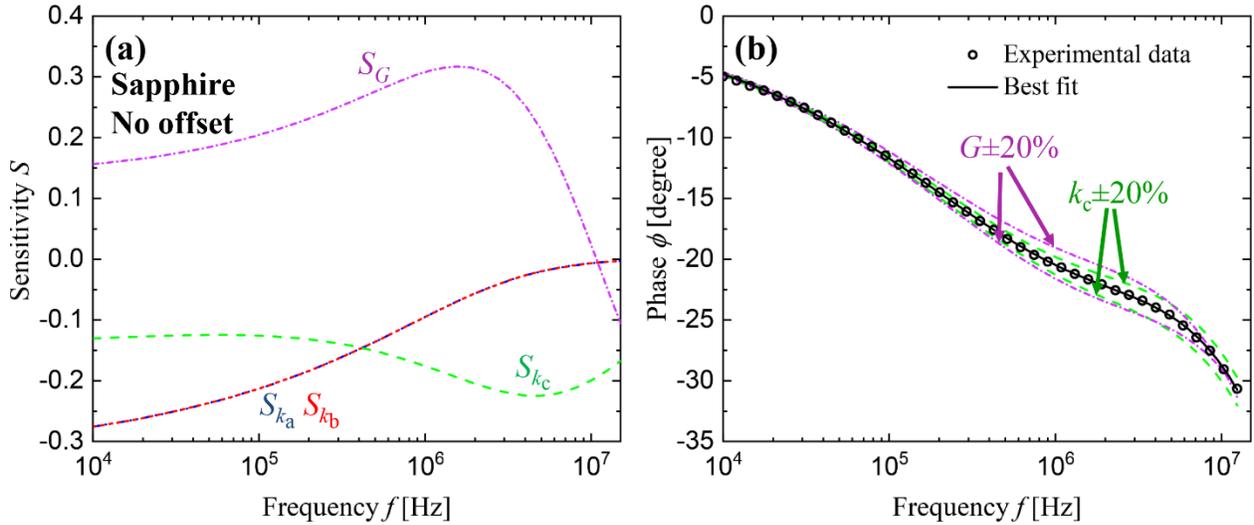

Figure 4. (a) Sensitivities of $\phi(f)$ to the thermal conductivity tensor elements and thermal interface conductance for sapphire with no beam offset. (b) Measured phase lag as a function of frequency for sapphire with no beam offset. The best fit curve and the results obtained by perturbing $G$ and $k_c$ by ±20% are also plotted.

The measured phase lag data for this zero beam offset case along with the best fit $\phi(f)$ curve are shown in Fig. 4b. To clearly visualize the sensitivity, the curves are also plotted when the $G$ and $k_c$ values from the best fit are varied by ±20%, holding all other parameters constant. To ensure the uniqueness of the $(G, k_c)$ fit, it was repeated using a wide range of initial guesses, and we found that the final fit converges consistently to the same property values independent of the initial guesses. We similarly confirmed that the rest of the fits shown later in the paper are also independent of the initial guesses. Consistent with the sensitivity plot of Fig. 4a, the difference between the best fit and ±20% perturbed $G$ and $k_c$ curves in Fig. 4b is significant at the intermediate and high frequencies, from ~ $5 \times 10^5$ to $6 \times 10^6$ Hz.

From the best fit in Fig. 4(b) the cross-plane thermal conductivity and thermal interface conductance are determined as summarized in Table 2. This measured $k_{zz}$ value is within 10% of a handbook value [35], and the measured $G$ is also reasonable [37]. Note that the thermal conductivity values in Table 2 are expressed in the xyz coordinate system, which for the sapphire sample is related to the crystallographic abc coordinate system as depicted in Fig. 3.

Table 2. Measured and literature values for the thermal conductivity tensor elements of all samples measured in this work, along with the measured thermal interface conductance. All values are for room temperature. The uncertainties are calculated based on Eq. (7) and Table 1. The relative orientation of each sample's xyz and abc coordinate systems are depicted in Fig. 3 (sapphire, HOPG), Fig. 10 (aligned x-cut quartz), and Fig. 14. For aligned x-cut quartz, the measured $k_{xx}$ and $k_{yy}$ are based on the data in Fig. 12(b) for $y_o = 5$ μm. Literature sources: sapphire [35], HOPG [38], and x-cut quartz [33].

| Sample | Literature | | | | Measured | | | | |
| --- | --- | --- | --- | --- | --- | --- | --- | --- | --- |
| | $k_{xx}$ W/m-K | $k_{yy}$ W/m-K | $k_{xy}$ W/m-K | $k_{zz}$ W/m-K | $k_{xx}$ W/m-K | $k_{yy}$ W/m-K | $k_{xy}$ W/m-K | $k_{zz}$ W/m-K | $G$ MW/m²-K |
| Sapphire | 30.3 | 30.3 | 0 | 32.5 | 36.3 ± 2.3 | 35.0 ± 1.7 | N/A | 35.7 ± 2.1 | 48.6 ± 1.1 |
| HOPG | 1900 ± 240 | 1900 ± 240 | 0 | 6.5 ± 0.7 | 1744 ± 185 | 1712 ± 145 | N/A | 5.9 ± 0.4 | 25.9 ± 1.1 |
| x-cut quartz (aligned) | 10.8 | 6.2 | 0 | 6.2 | 10.5 ± 1.2 | 6.2 ± 0.6 | N/A | 6.1 ± 0.4 | 45.8 ± 1.1 |

| x-cut quartz (Not aligned) | 7.4 | 9.7 | -2.0 | 6.2 | 7.7 ± 1.1 | 8.8 ± 1.1 | -2.4 ± 0.2 | 6.1 ± 0.4 | 45.8 ± 1.1 |

The uncertainties of the measured $k_c$ (= $k_{zz}$) and $G$ given in Table 2 are determined based on Eq. (7), which is evaluated numerically following Ref. [29]. Briefly, the following three step process is used: (1) Determine the $k_c$ and $G$ values by fitting the modeled $\phi(f)$ to experimental $\phi(f)$ based on the nominal values of the input parameters $\beta$ from Table 1; (2) Repeat Step 1 by varying one input parameter $\beta$ by its uncertainty $\pm\Delta\beta$ to determine the quantities $\pm\Delta k_{c\beta}$ and $\pm\Delta G_\beta$ (i.e., the difference between the best-fit values obtained from Steps 1 and 2); (3) Repeat Steps 1 and 2 to determine all $\Delta k_{c\beta}$ and $\Delta G_\beta$ and use Eq. (7) to calculate the final uncertainties of the measurements of $k_c$ and $G$. This same method is utilized for all uncertainty calculations in the rest of the paper. Due to their relatively high sensitivity values, the uncertainties of the obtained $k_c$ and $G$ for sapphire are relatively small, less than ±6%, as seen in Table 2. Furthermore, as $k_c$ and $G$ are most sensitive at the intermediate and high frequencies, the largest contribution to their uncertainties come from $C_{Au}$ and $t_{Au}$ because these quantities also have the largest $S$ in that frequency regime, as shown in Fig. 2(a) and discussed in Section 3.3. Note that the fit in Fig. 4b actually used four parameters ($k_a$, $k_b$, $k_c$, and $G$), but as explained several paragraphs above the $k_a$ and $k_b$ results are discarded because of the impossibility of decoupling them from each other. (It is still interesting to note that the geometric mean of their products, $\sqrt{k_a k_b} \approx 35$ W/m-K, remains reasonably consistent with the literature values. This is expected because the joint confidence interval of $k_a$ and $k_b$ for small beam offsets has a generally hyperbolic shape, as will be explained later in this Section around Fig. 7).

Next, the two in-plane thermal conductivities of sapphire will be measured independently of each other by adding a beam offset. We first use sensitivity analysis to determine the suitable frequency and beam-offset regimes for this measurement. Figure 5 shows the sensitivities of $\phi(f)$ to $k_a$ and $k_b$ for different offset distances in the y-axis direction. For increasing $y_o$, the sensitivities to the in-plane thermal conductivities become slightly smaller in magnitude but, crucially, much more distinguishable. This is expected since displacing the probe along the $y$ direction breaks the in-plane symmetry of the measurement, causing the measured $\Delta T$ response to now depend differently on the heat diffusion in $x$ vs. $y$ directions. Figure 5 also shows that the sensitivities to $k_a$ and $k_b$ are relatively high at the lower frequencies, from ~$10^4$ to $10^6$ Hz.

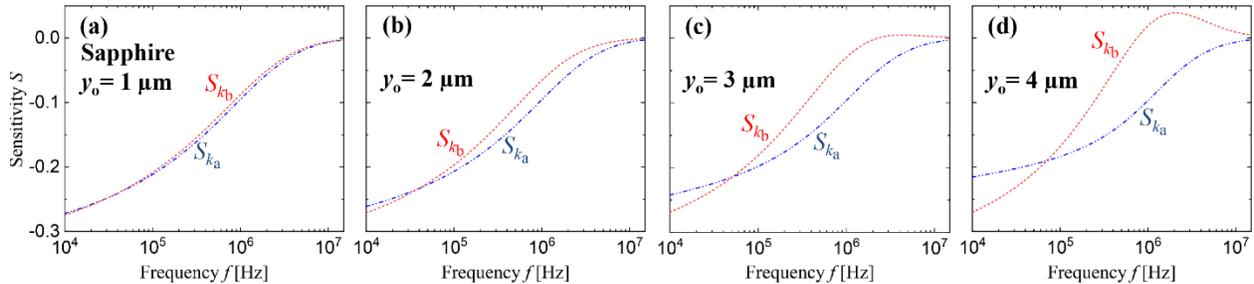

Figure 5. Sensitivities of $\phi(f)$ to $k_a$ and $k_b$ for sapphire with different beam offsets in the y-axis direction. As $y_o$ increases, the effects of $k_a$ and $k_b$ become increasingly distinguishable.

Now taking $G$ and $k_c$ as determined from Fig. 4b as known input parameters, for each offset distance the in-plane thermal conductivities $k_a$ and $k_b$ are obtained by fitting the measured phase data to the analytical model for that low- and intermediate-frequency range ($10^4$ to $10^6$ Hz). The measured phase lag data along with the best fits are shown in Fig. 6. The best fit values for $k_a$ and $k_b$ are given in Table 2 as obtained from the 4 μm beam-offset data, since at this offset we are best able to distinguish between $k_a$ and $k_b$, as seen from the sensitivity calculations of Fig. 5.

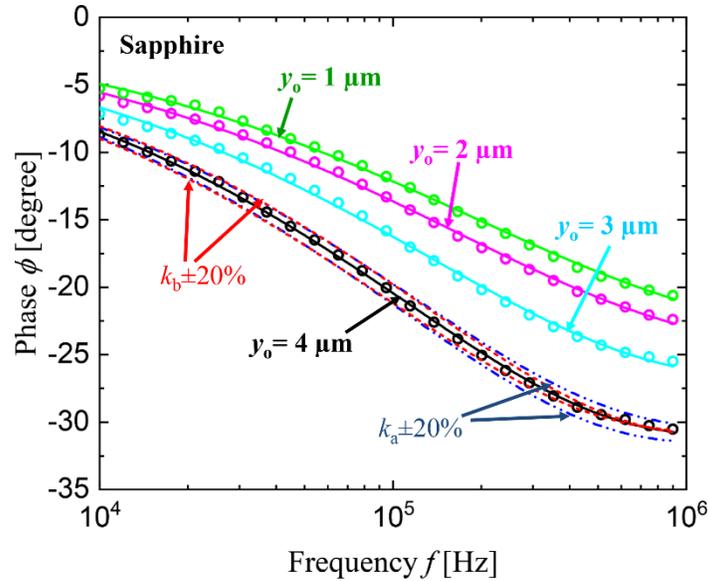

Figure 6. Measured phase lag (points) for sapphire along with the best fits (lines) for different beam offsets in the *y* direction. For these fits $G$ and $k_c$ are treated as constant, using the values determined in Fig. 4. The model results obtained by perturbing $k_a$ and $k_b$ by ±20% are also plotted for 4 μm beam offset.

In order to more clearly visualize how the beam offset distance influences the fit $k_a$ and $k_b$ results, in Fig. 7 we also consider the shape and size of the "confidence interval valley" in $k_a$-$k_b$ space (similar multiparameter confidence contour representations have been used elsewhere [19], [21], [23]). Specifically, for each value of $y_o$, we vary $k_a$ and $k_b$ continuously and for every ($k_a$, $k_b$) pair calculate the root-mean-square (rms) residual, $r$, of the model $\phi(y_o, f)$ as compared to the measured $\phi(y_o, f)$ data in Fig. 6. Then to present all results on a similar scale, for each $y_o$ we normalize the resulting $r(k_a, k_b)$ surface by the minimum $r$ found from that ($k_a$, $k_b$) sweep. We call this the normalized residual, $R$.

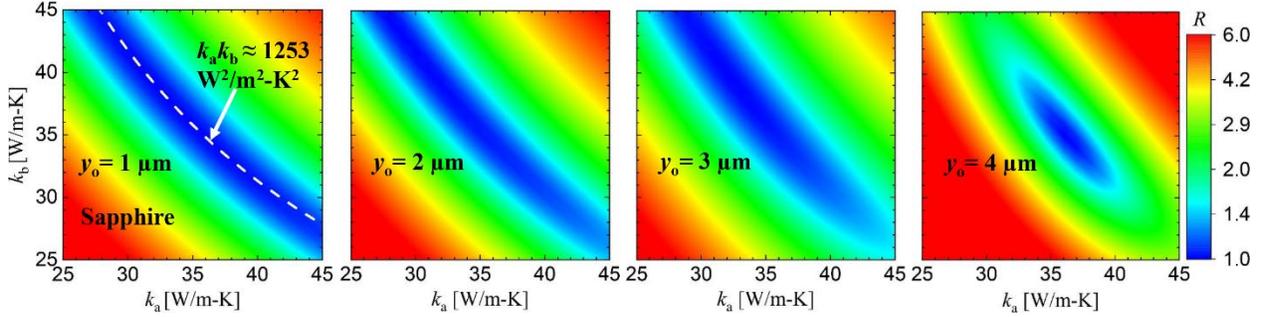

Figure 7. Normalized rms residual, $R$, as a function of $k_a$ and $k_b$ for different beam offsets. Here $R$ is calculated separately for each panel by normalizing by each panel's minimum residual. The zone of uncertainty (valley of small $R$) clearly becomes tighter for larger $y_o$. The dashed line in the leftmost panel denotes the hyperbola curve demonstrating the constant value of $k_a k_b$ obtained from the best fit in Fig. 6.

The resulting $R(y_o, k_a, k_b)$ are shown in Fig. 7. For all $y_o$ the small-$R$ confidence valley (blue region) has a generally hyperbolic shape, $k_a k_b \approx 1253$ W$^2$/m$^2$-K$^2$ = (35.4 W/m-K)$^2$. As such, their harmonic mean, $\sqrt{k_a k_b}$, can be determined from any of the $y_o$ datasets. (A similar hyperbolic shaped confidence valley is observed in $R(k_a, k_b)$ when there is no beam offset (results omitted here for brevity), which explains why $\sqrt{k_a k_b} \approx 35$ W/m-K obtained from the best fit in Fig. 4b is reasonable, as mentioned above.) However, for small and intermediate offsets ($y_o \leq 3$ μm) this hyperbolic valley extends very far in the diagonal direction from upper-left to lower-right in their respective panels of Fig. 7, which means that $k_a$ and $k_b$ cannot be untangled from each other. The hyperbolic valley gradually becomes tighter with increasing $y_o$, such that $k_a$ and $k_b$ can each be determined individually with reasonably uncertainty using a large offset distance of 4 µm. Larger offsets ($y_o$ > 4 µm) were not pursued here since as seen in Fig. 5 the absolute magnitude of the sensitivities falls off for larger $y_o$, leading to increased uncertainties again.

Focusing on the $y_o$ = 4 µm case, the best fit values are $k_a$ = 36.3 ± 2.3 W/m-K and $k_b$ = 35.0 ± 1.7 W/m-K. These values are consistent with literature values. Importantly, unlike prior BO-FDTR studies here the measurement allowed for the possibility of $k_a \neq k_b$ yet still found $k_a \approx k_b$ to within 4% which is better than the experimental uncertainty. This demonstrates how the present scheme correctly recovers the required transversely isotropic **k** tensor for the sapphire orientation considered in Fig. 3.

The uncertainties of the measured $k_a$ and $k_b$ are determined based on the three step procedure described above for calculating the uncertainties of $k_c$ and $G$. Regarding the treatment of $k_c$ and $G$ while calculating $\Delta k_a$ and $\Delta k_b$, we take into account the fact that $k_c$ and $G$ are intermediate quantities rather than independent inputs like those listed in Table 1. Therefore, when perturbing each input parameter $\beta$ by its uncertainty $\Delta \beta$ (for example perturbing $k_{Au}$ by 2%), we re-fit the zero-beam-offset $\phi(f)$ data of Fig. 4b to obtain correspondingly perturbed best-fit values for $k_c$ and $G$, and then use these perturbed $\beta$, $k_c$, and $G$ when re-fitting the 4 µm offset data of Fig. 6 to get the contributions $\Delta k_{a\beta}$ and $\Delta k_{b\beta}$ for use in Eq. 7. The final uncertainty in $k_a$ and $k_b$ mostly stems from the sapphire heat capacity and the pump radius, because as seen in Fig. 2(b) $|S_{C_{Sapphire}}|$ and $|S_{w_0}|$ are large at the same low and intermediate frequency range which is used to fit $\phi(f)$ for $k_a$ and $k_b$. It is

also worth noting that there should be no difference in the measurements by adding the beam offset in either the x- or y-axis directions for materials with in-plane thermal conductivity isotropy, like sapphire in the orientation used here.

*5.1.2 HOPG*

Similar measurements were performed on HOPG, again in the configuration of Fig. 3. Unlike sapphire, HOPG has an order of magnitude contrast between the in-plane and cross-plane thermal conductivities, with $k_a = k_b \gg k_c$. The procedure for HOPG thermal conductivity measurements is similar to that used for sapphire. When there is no beam offset, $\phi(f)$ is sensitive to the cross-plane thermal conductivity and thermal interface conductance at high frequencies, from $\sim 10^6$ to $1.5 \times 10^7$ Hz (see Fig. 8a). Just as with sapphire, due to crystal symmetry $S_{k_a} = S_{k_b}$ in this configuration and so we focus only on the fits for $k_c$ and $G$. The measured phase lag data for this frequency range with the best fit curve and effects of perturbing $k_c$ and $G$ are shown in Fig. 8b. The resulting $k_c$ and $G$ were found to be $5.9 \pm 0.4$ W/m-K and $25.9 \pm 1.1$ MW/m²-K, respectively.

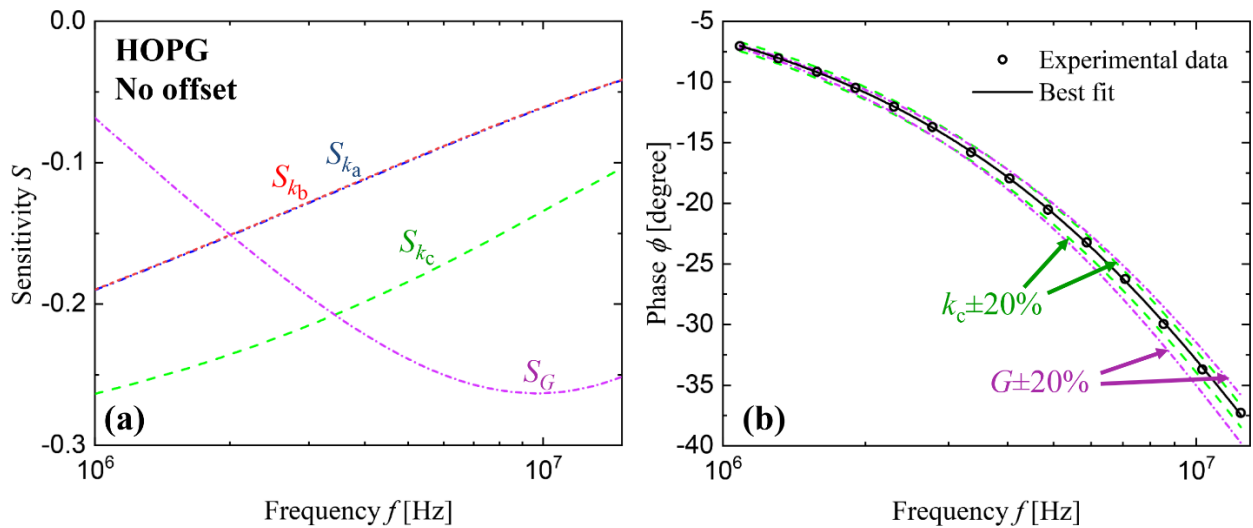

Figure 8. (a) Sensitivities of $\phi(f)$ to the principal thermal conductivities and thermal interface conductance for HOPG with no beam offset. (b) Measured phase lag data as a function of frequency for HOPG with no beam offset. The best fit curve and the effects of perturbing $G$ and $k_c$ by ±20% are also plotted.

Adding a beam offset of $y_o = 5$ μm to the HOPG measurements, the sensitivities to $k_a$ and $k_b$ become readily distinguishable, especially at lower frequencies (see Fig. 9a). Therefore, the phase lags were measured for $f$ from $10^5$ to $6 \times 10^6$ Hz with 5 μm beam offset. Using the $k_c$ and $G$ values determined from Fig. 8b, fitting Fig. 9b for the in-plane thermal conductivities gives $k_a$ and $k_b$ respectively of 1744 ± 185 and 1712 ± 145 W/m-K. Thus, the measured $k_a$ and $k_b$ are nearly identical, being within 2% of each other, while being quite different from the cross-plane thermal conductivity by a factor of nearly 300. This is all as expected for a transversely isotropic material like HOPG. The comparison between our HOPG measurements and literature values in Table 2 shows all three principal conductivities have good agreement. Similar to the sapphire case, extended sensitivity analysis analogous to Fig. 2 (details omitted here for brevity) for HOPG show that the largest uncertainty for HOPG's $k_c$ and $G$ comes from the heat capacity of the Au transducer layer, while the largest uncertainty for HOPG's in-plane thermal conductivities is from the pump radius.

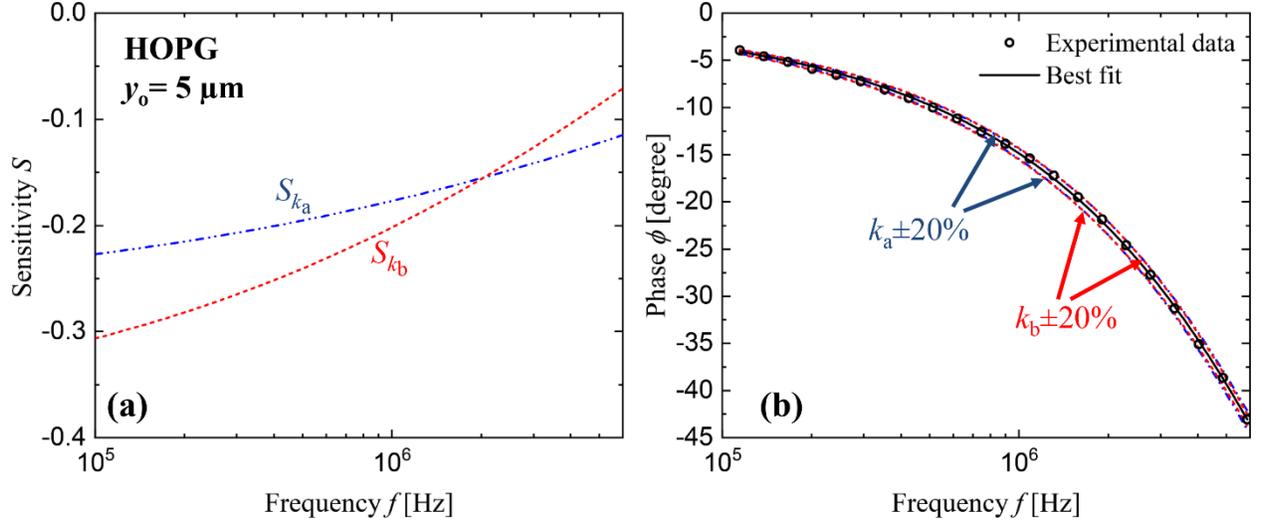

Figure 9. (a) Sensitivities of in-plane thermal conductivities for HOPG with 5 μm beam offset in the *y* direction. (b) Measured phase lag data for HOPG with 5 μm beam offset in the *y* direction. The best fitting curve and the results obtained by varying $k_a$ and $k_b$ by ±20% are also plotted. For these fits $G$ and $k_c$ are treated as constant, using the values determined in Fig. 8.

5.2 Thermal conductivity measurements for a transversely anisotropic material

*5.2.1 Case 1: Known orientation of the crystal principal axes*

The main advantage of the current method compared to the previous BO-FDTR studies [16], [17], [23] is the ability to obtain the in-plane thermal conductivities individually. To demonstrate this we now present measurements on "x-cut" quartz (<110> α-SiO$_2$), which has $k_a = k_b < k_c$ [33]. The x-cut quartz is first well aligned as depicted in Fig. 10, such that the principal in-plane thermal conductivities, $k_c$ and $k_b$, and cross-plane thermal conductivity, $k_a$, are respectively along the x-, y- and z-axis directions. Thus in this orientation $k_{xx} > k_{yy} = k_{zz}$, giving an interesting in-plane anisotropy. Also, all off-diagonal terms of the **k** tensor vanish for this orientation.

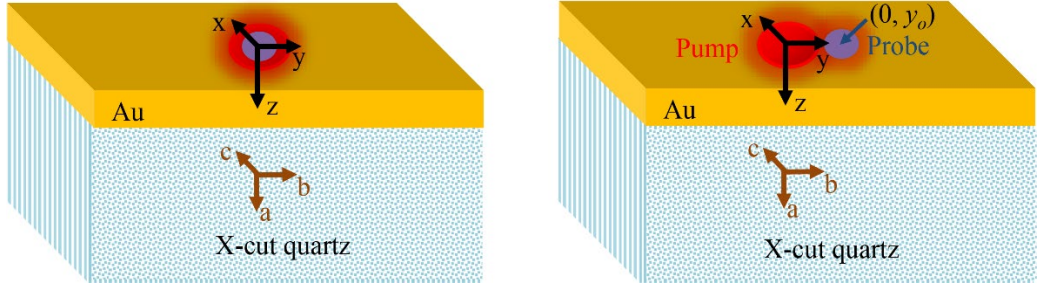

Figure 10. Schematic for the well-aligned x-cut quartz sample with co-aligned (left) and offset (right) beams. The x, y, and z-axes are along the principal c-, b-, and a-axis directions, respectively.

Without beam offset, the sensitivities of $\phi(f)$ to the in-plane thermal conductivities of this quartz follow a very similar trend as each other, as depicted in Fig. 11(a), so it is difficult to decouple $k_b$ and $k_c$ by fitting a measurement. On the other hand, Fig. 11(a) also shows that the thermal interface conductance and cross-plane thermal conductivity $k_a$ can be decoupled and have relatively large sensitivity values at high frequencies. These trends are all similar to phenomena described above for sapphire and HOPG when there is no beam offset. Fig. 11(b) shows the measured phase lag data along with the best fit. After fitting, the resulting $k_a$ (same as $k_{zz}$) and $G$ values are summarized in Table 2.

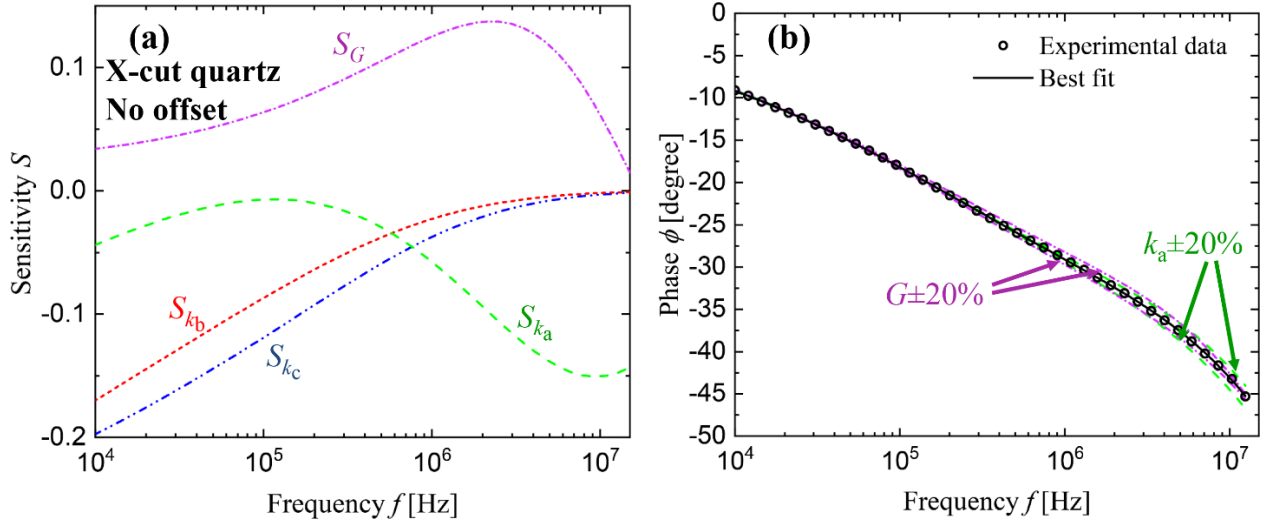

Figure 11. (a) Sensitivities of $\phi(f)$ to the principal thermal conductivities and thermal interface conductance for x-cut quartz with no beam offset, aligned as in Fig. 10. (b) Measured phase lag as a function of frequency. The best fitting curve and the results obtained by varying $G$ and $k_a$ by ±20% are also plotted.

Next the in-plane thermal conductivities can be individually obtained by adding a beam offset in either x- or y-axis directions. Note that unlike the sapphire and HOPG measurements presented above, for this orientation of x-cut quartz the two in-plane directions (x,y) are *not* equivalent, because here $k_{xx} \neq k_{yy}$ (since $k_c \neq k_b$ in Fig. 10). Therefore here it is meaningful to consider offsets in both y and x separately, and this is done in Figs. 12(a) and 13(a), respectively. When the beam-offset distance is 5 μm in either x- or y- directions, the sensitivities of the in-plane thermal conductivities become quite distinct from each other and have usefully large values at frequencies from $10^3$ to ~$10^6$ Hz, as shown in Figs. 12(a) and 13(a). (For comparison, recall from Fig. 11(a) that $S_{k_b}$ and $S_{k_c}$ are nearly indistinguishable when the offset is zero.)

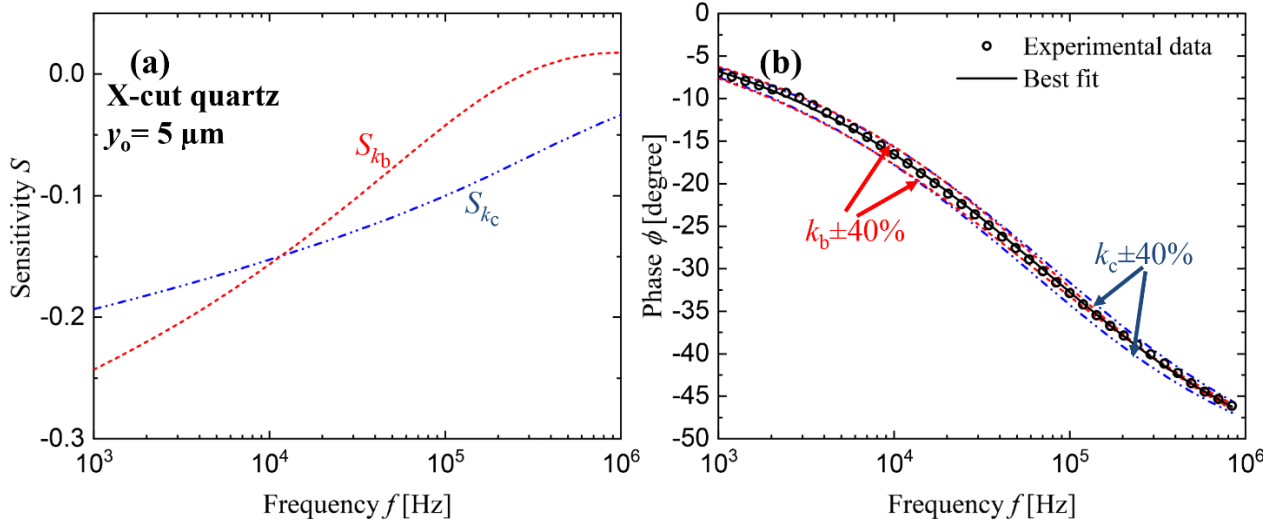

Figure 12. (a) Sensitivities of $\phi(f)$ to the in-plane thermal conductivities for x-cut quartz with 5 μm beam offset in the y-axis direction. (b) Measured phase lag as a function of frequency. The best fit curve and the results obtained by varying $k_b$ and $k_c$ by ±40% are also plotted. For these fits $G$ and $k_a$ are treated as constant, using the values determined in Fig. 11.

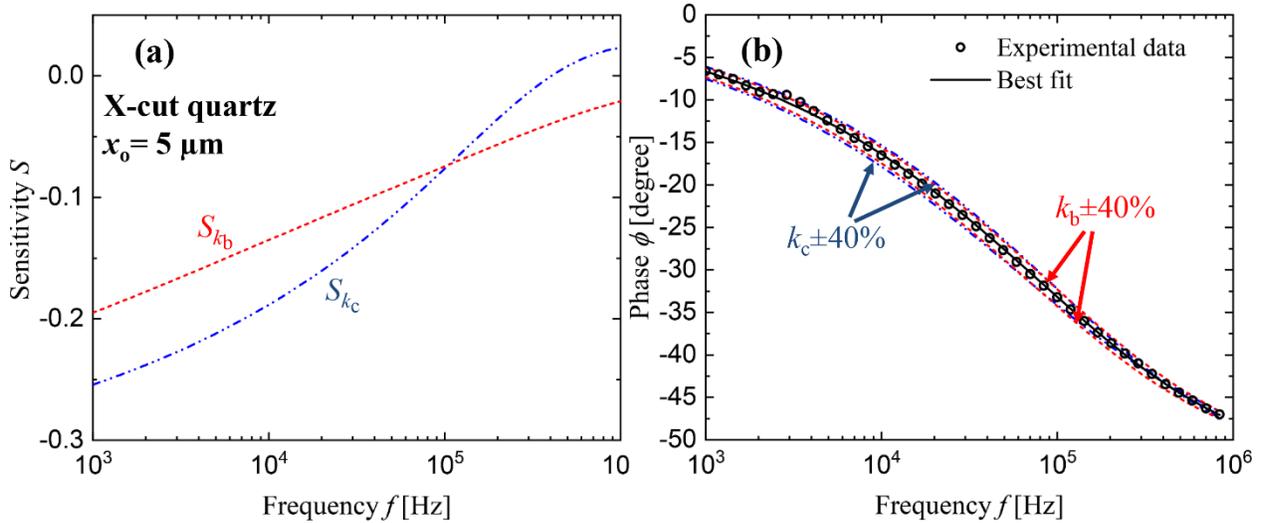

Figure 13. Similar to Fig. 12 but with the beam offset now in the x direction. (a) Sensitivities of $\phi(f)$ to the in-plane thermal conductivities for x-cut quartz with 5 μm beam offset in the x-axis direction. (b) Measured phase lag as a function of frequency. The best fit curve and the results

obtained by varying $k_b$ and $k_c$ by ±40% are also plotted. For these fits $G$ and $k_a$ are treated as constant, using the values determined in Fig. 11.

By fitting the measured phase lag data for this frequency range at 5 μm beam offset in the x- and y-axis directions to the theoretical model (see Figs. 12(b) and 13(b)), similar in-plane thermal conductivity results are obtained from these two different datasets, with good self-consistent agreement of better than 5%. Specifically, for the y-offset case we find $k_c = 10.5 \pm 1.2$ W/m-K and $k_b = 6.2 \pm 0.6$ W/m-K, and for the x-offset case the fitting yields $k_c = 10.0 \pm 1.3$ W/m-K and $k_b = 6.3 \pm 1.3$ W/m-K. Furthermore, the uncertainty of $k_b$ becomes larger (smaller) when offsetting the beam along x-axis (y-axis) direction due to smaller (larger) sensitivity values $S_{k_b}$ as depicted in Figs. 12(a) and 13(a). Also, the relative uncertainties ($\Delta k/k$) of these in-plane thermal conductivities are larger than the case of previous experiments on sapphire because quartz has smaller thermal conductivity overall.

Although the uncertainties of the in-plane thermal conductivities of quartz are still dominated by the pump radius, the uncertainties related to the Au transducer layer ($\Delta k_{Au}$, $\Delta C_{Au}$, and $\Delta t_{Au}$) also play an important role here due to the more pronounced lateral heat spreading in the Au transducer, as compared to the previous experiments on sapphire with its higher thermal conductivity. Further improvement regarding the uncertainty could be realized by shifting to a lower thermal conductivity transducer [20]. Overall, all three measured thermal conductivities for quartz agree quite well with literature values (see Table 2). Also, we emphasize that the extracted $k_a$ and $k_b$ (represented in Table 2 by $k_{zz}$ and $k_{yy}$, respectively) were determined independently in our measurements yet are nearly identical (to better than 2%), as required for this crystal structure and orientation in Fig. 10.

*5.2.2 Case 2: Arbitrary in-plane alignment*

As a final experimental demonstration, we consider the more general case where the in-plane orientation of the crystal principal axes is arbitrary and unknown. In this case the off-diagonal element $k_{xy}$ of the thermal conductivity tensor is now finite, and must be determined. Here we demonstrate this measurement capability using an x-cut quartz sample that has been intentionally rotated such that the principal c-axis direction is now $\psi = 30°$ away from the y-axis direction, as shown in Fig. 14(a). From elementary tensor identities [39] it is known that $k_{xx} = k_b \cos^2\psi + k_c \sin^2\psi$, $k_{yy} = k_c \cos^2\psi + k_b \sin^2\psi$, and $k_{xy} = (k_b - k_c)\cos\psi \sin\psi$. Note that the z-axis is still parallel to the crystal a-axis in this configuration, so that $k_{zz} = k_a$ and the other two off-diagonal terms, $k_{xz}$ and $k_{yz}$, are still zero. Thus in this alignment there are four unknown parameters of the **k** tensor: $k_{xx}$, $k_{yy}$, $k_{zz}$, and $k_{xy}$.

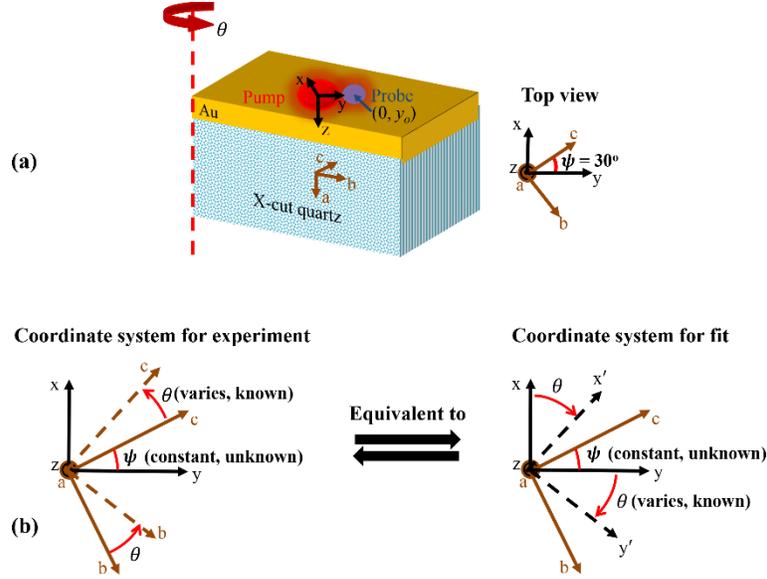

Figure 14. (a) Schematic of x-cut quartz when the principal c-axis is rotated $\psi = 30°$ away from the y-axis, representing arbitrary in-plane anisotropy including finite $k_{xy}$. The z-axis and crystal a-axis remain parallel. (b) The sample is rotated counterclockwise further by additional various known angles $\theta$. The two schematics show coordinate systems used for experiment (left: applying $\theta$ to the abc axes) and model fit (right: applying $-\theta$ to the xyz axes). The systems are equivalent because all relative angles are the same. The $\theta$ are varied and known while $\psi$ is fixed and is treated as an unknown parameter when analyzing the experiments.

As already shown above, $k_{zz}$ and $G$ are readily obtained through coincident-spot measurements which are independent of $\psi$, and thus we proceed using the values obtained in the previous experiments of Section 5.2.1. For the three remaining unknowns $k_{xx}$, $k_{yy}$, and $k_{xy}$, in principle it may seem obvious to fit $\phi(f)$ data for $y_o = 5$ μm beam offset and thus obtain these three parameters simultaneously. However, sensitivity analysis shows this to be impractical here: as shown in Fig. 15(a), $k_{xx}$ and $k_{yy}$ can still be individually determined due to their reasonably large and distinct sensitivities, but unfortunately in this case the sensitivity for $k_{xy}$ is very small, such that the $k_{xy}$ result from fitting will not be accurate.

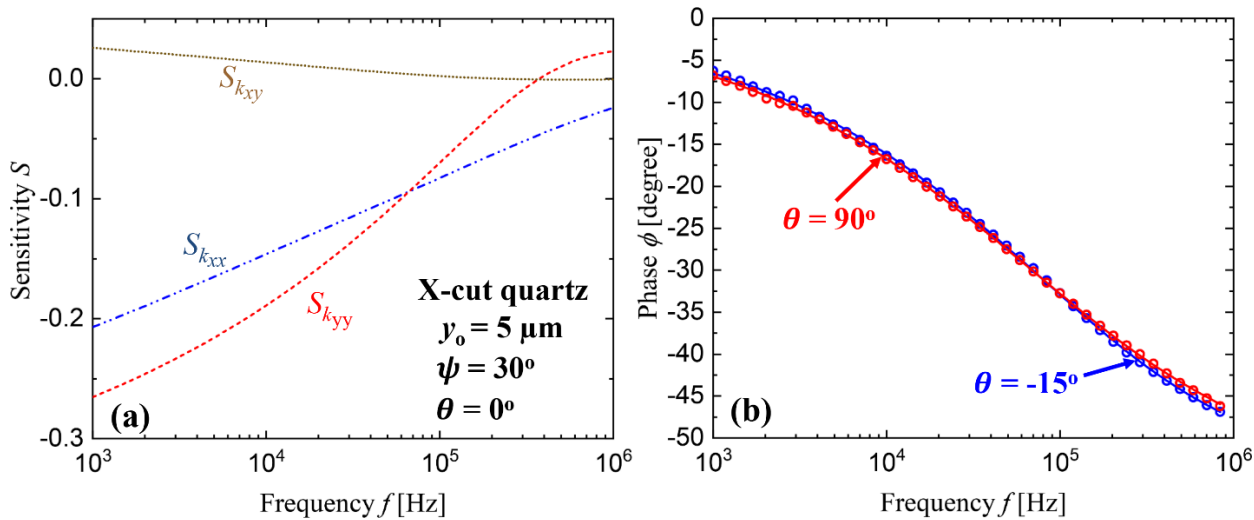

Figure 15. (a) Sensitivities of the in-plane thermal conductivity tensor elements at 5 μm beam offset in the y direction, for x-cut quartz as in Fig. 14 with $\psi = 30°$ and $\theta = 0°$. (b) Measured phase lag as a function of frequency at 5 μm beam offset in the y direction for $\theta = -15°$ and 90°. (Results equivalent to $\theta = 60°$ and -30° are given in Figs. 12(b) and 13(b).)

Instead we introduce an additional measurement principle to extract all three of $k_{xx}$, $k_{yy}$, and $k_{xy}$. Specifically, the sample is rotated counterclockwise about the a-axis (equivalent to the z-axis) by various known angles $\theta$. Thus the total angle between the c-axis and y-axis is $\theta + \psi$, where $\psi$ is treated as unknown for the purposes of this measurement demonstration (See Fig. 14(b), left). For simplicity in the analysis to come, here we consider a new coordinate system x′y′ rotated clockwise by an angle $\theta$ with respect to the original coordinate xy as depicted in Fig. 14(b, right). This clockwise xy to x′y′ coordinate rotation is equivalent to the counterclockwise sample rotation as done in the experiments. Then for each new orientation $k_{x'x'}$ and $k_{y'y'}$ are measured, and finally the resulting $k_{x'x'}(\theta)$ and $k_{y'y'}(\theta)$ data will be used to determine the full in-plane thermal conductivity tensor for the reference configuration ($\theta = 0$), as detailed next.

Representative measured phase lag data for 5 μm beam offset in the y-axis direction for two different rotation angles of $\theta$ = -15° and 90° are shown in Fig. 15(b). Also, note that the phase lag data at $y_o$ = 5 μm for the rotation angles of $\theta$ = 60° and -30° are equivalent to the results already shown in Figs. 12(b) and 13(b), respectively, since for example with $\theta$ = 60° and $\psi$ = 30°, from Fig. 14(b) we see the y-axis becomes aligned with the b-axis, corresponding to the configuration already measured in Fig. 12(b). In this way, we obtain $\phi(f)$ data for all $\theta$ from -30° to 105° in 15° steps, always holding $y_o$ constant at 5 μm. For each of these $\theta$, we fit the corresponding $\phi(f)$ dataset with the three parameters, $k_{x'x'}$, $k_{y'y'}$ and $k_{x'y'}$. The resulting $k_{x'x'}(\theta)$ and $k_{y'y'}(\theta - 90°)$ are shown in Fig. 16, while the $k_{x'y'}(\theta)$ fit results are discarded due to their small sensitivity values (recall Fig. 15(a)). Note that the error bars of the $k_{y'y'}$ measurements are smaller than those of the $k_{x'x'}$ measurements, since the beam offset is in the y-axis direction.

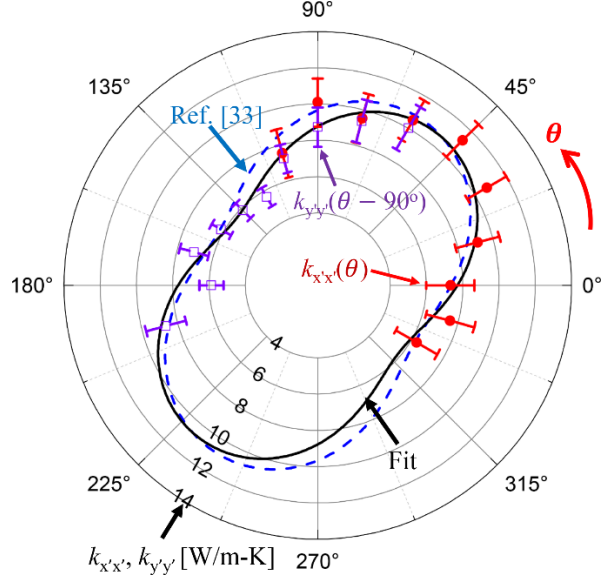

Figure 16. Measured thermal conductivities (points) for various in-plane sample rotations $\theta$, for x-cut quartz with $\psi = 30°$ in the configuration depicted in Fig. 14. Red and purple points are the measurements of $k_{x'x'}(\theta)$ and $k_{y'y'}(\theta - 90°)$, respectively. The solid black line is a fit of $k_{x'x'}(\theta)$ according to Eqs. (8) to all 20 points. The dashed blue line represents the expected $k_{x'x'}(\theta)$ based on the principal **k** values from Ref. [33] and applying a $\psi = 30°$ rotation.

The in-plane thermal conductivity tensor in this x'y' coordinate system, $\mathbf{k'_{in\text{-}plane}} = \begin{bmatrix} k_{x'x'} & k_{x'y'} \\ k_{y'x'} & k_{y'y'} \end{bmatrix}$, is related to that in the original coordinate system, $\mathbf{k_{in\text{-}plane}} = \begin{bmatrix} k_{xx} & k_{xy} \\ k_{yx} & k_{yy} \end{bmatrix}$, by a well-known tensor identity $\mathbf{k'_{in\text{-}plane}} = \mathbf{T}^{-1} \mathbf{k_{in\text{-}plane}} \mathbf{T}$, where $\mathbf{T} = \begin{bmatrix} \cos\theta & \sin\theta \\ -\sin\theta & \cos\theta \end{bmatrix}$ is the rotation matrix [39]. In other words, one can calculate $k_{x'x'}$ as a function of rotation angle as

$$k_{x'x'} = k_{xx}\cos^2(\theta) + k_{yy}\sin^2(\theta) - 2k_{xy}\cos(\theta)\sin(\theta),  \tag{8a}$$

and similarly,

$$k_{y'y'} = k_{yy}\cos^2(\theta) + k_{xx}\sin^2(\theta) + 2k_{xy}\cos(\theta)\sin(\theta),  \qquad (8b)$$

where it is helpful to recognize that

$$k_{y'y'}(\theta \pm 90^\circ) = k_{x'x'}(\theta) \qquad (8c)$$

Therefore, $k_{xx}$, $k_{yy}$ and $k_{xy}$ of quartz in the original orientation considered in Fig. 14 ($\psi = 30°$, $\theta = 0°$) are obtained by using Eqs. (8) to fit the $\theta$-dependent $k_{x'x'}(\theta)$ and $k_{y'y'}(\theta)$ shown by the points in Fig. 16, and the best fit is the "peanut-shaped" curve also shown in Fig. 16. The corresponding fit values of $k_{xx}$, $k_{yy}$, and $k_{xy}$ are given in Table 2. Subsequently, the angle between the c- and y-axis can be determined from these measured in-plane thermal conductivity elements via $\psi = \frac{1}{2}\tan^{-1}(2k_{xy}/(k_{xx}-k_{yy}))$, giving $\psi_{fit} = 38.5 \pm 5.5°$, which is close to the expected value of $\psi_{exact} = 30°$. The difference might be due to some errors of the rotation angles introduced by the rotation stage and an imperfect initial sample alignment along the principal axes.

### 5.3 A numerical experiment demonstrating the potential to fit all three in-plane tensor elements from a single sample orientation

As noted above in the discussion of Fig. 15(a), for the case of x-cut quartz it was infeasible to determine all three of $k_{xx}$, $k_{yy}$, and $k_{xy}$ from a single sample orientation, because the sensitivity to $k_{xy}$ was too low. However, for a material with higher in-plane thermal conductivity anisotropy such a measurement becomes possible. To demonstrate this, here we present a case study using a fictitious sample material. The material is oriented like in Fig. 14 with $k_c = 5$ W/m-K, $k_b = 50$ W/m-K, and $k_a = 50$ W/m-K. As such, the in-plane anisotropy ratio, $k_b/k_c$, is high with a value of 10. Taking into account the $\psi = 30°$ rotation, in the xyz coordinate system the thermal conductivity tensor is

$$\mathbf{k} = \begin{bmatrix} 38.8 & 19.5 & 0 \\ 19.5 & 16.3 & 0 \\ 0 & 0 & 50 \end{bmatrix} \text{W/m-K} \tag{9}$$

Now following the same procedure as the experiments on quartz shown in Section 5.2.2, we assume that $G$ and $k_{zz}$ (= $k_c$) are known from a separate measurement with aligned pump and probe spots. We assume $C_{\text{sample}} = 2.0$ MJ/m³-K and $G = 40$ MW/m²-K, and all other experimental input parameters with their uncertainties are taken from Table 1. Thus the unknown parameters of interest are the in-plane thermal conductivity elements, $k_{xx}$, $k_{yy}$, and $k_{xy}$. These three in-plane thermal conductivity elements can now be accurately and simultaneously determined by simply using a single $\phi(f)$ dataset at $y_o = 5$ µm, since they have quite high and distinct sensitivity values as shown in Fig. 17(a).

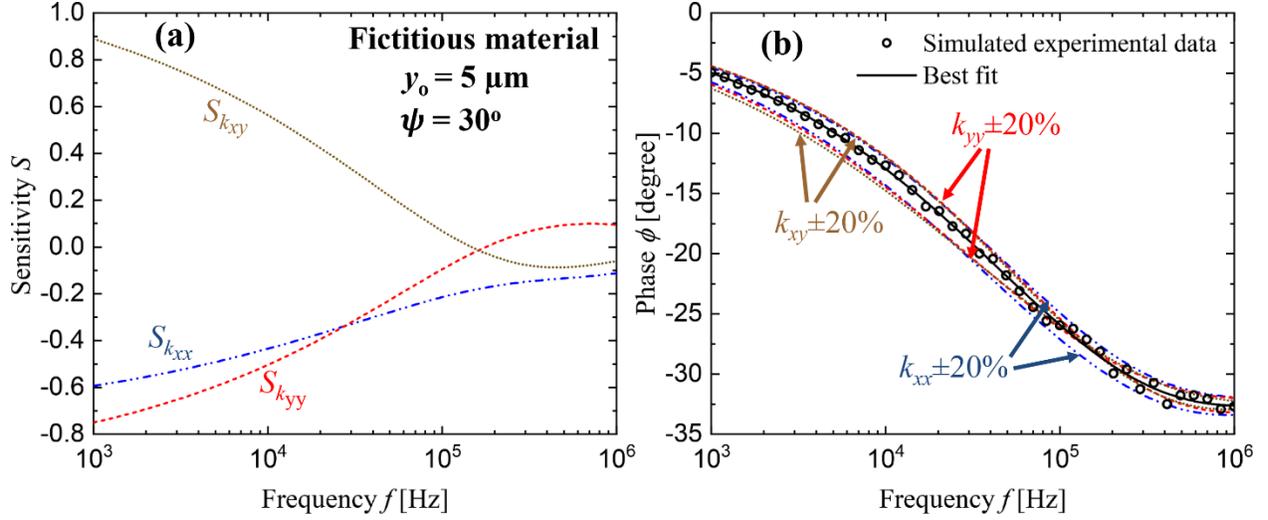

Figure 17. (a) Sensitivities of $\phi(f)$ to the in-plane thermal conductivity tensor elements for the fictitious material described in Eq. (9) in the configuration of Fig. 14, with 5 μm beam offset in the y direction. (b) Simulated experimental phase lag data including noise (points), along with the best fit curve (line) for this fictitious material. The results obtained by varying $k_{xx}$, $k_{yy}$, and $k_{xy}$ by ±20% are also plotted.

We will demonstrate how these three unknown parameters can be determined by simulating experiments, as follows. First, the analytical phase lag results corresponding to 40 frequencies spread over the range from $10^3$ to $10^6$ Hz at $y_o = 5$ μm are calculated using the exact **k** tensor given in Eq. (9). Then, experimental noise is simulated by adding random $\phi$ uncertainty ranging from -3% to +3% (uniformly distributed) to each data point. As such, the final simulated data is obtained and depicted by the points in Fig. 17(b). Next, the three in-plane thermal conductivity elements are determined by a simultaneous three-parameter fit of the analytical model to the simulated experimental data, as shown by the line in Fig. 17(b). The resulting fit $k_{xx}$, $k_{yy}$, and $k_{xy}$ of the fictitious material in this orientation are respectively 39.0 ± 2.0, 15.1 ± 1.2, and 18.3 ± 1.1 W/m-K, which are in good agreement (errors are less than 8%) with the exact values from Eq. (9). As with all multiparameter fits in this paper, the fit parameter values were confirmed to converge uniquely regardless of the initial guesses. Based on these fitted results, the angle between the c- and y-axis is calculated as $\psi_{fit} = 28.4° \pm 2.3°$, which is quite close to the exact value used, $\psi_{exact} = 30°$. Therefore, for highly anisotropic materials like this fictitious material, these numerical experiments demonstrate how the arbitrary in-plane thermal conductivity elements and principal in-plane directions can be determined by performing only a single beam-offset experiment. Such measurements are simpler since sample rotation is not required, in contrast to the example of the quartz experiments shown in Fig. 16. Either approach relies on having one additional measurement as well, with beams co-aligned, to obtain $G$ and $k_{zz}$.

## 6. Summary

We extend beam-offset FDTR from its previous application for transversely isotropic materials [16, 17, 23] to the more general case of transversely anisotropic materials, that is, materials lacking in-plane symmetry. The thermal conductivity tensor elements of three different materials spanning a wide range of thermal conductivity values (from ~ 6 to 2000 W/m-K) and degrees of anisotropy (from ~ 1 to 300) have been measured. To demonstrate its applicability for materials with arbitrary in-plane anisotropy, the new technique was used to measure the full in-plane thermal conductivity tensor of a transversely anisotropic material, x-cut quartz (<110> α-SiO$_2$), by combining a series of measurements in different directions. Extensive sensitivity analysis was used to determine the appropriate range of heating frequencies and beam offsets for all measurements. In addition, we performed a purely computational case study to show that the arbitrary in-plane thermal conductivity tensor of a fictitious material with a high in-plane anisotropy ratio can in principle be obtained from only a single sample orientation, rather than multiple sample orientations as was required for the x-cut quartz. While allowing for anisotropy, all measurements demonstrated excellent self-consistency in correctly identifying isotropic directions when present, with errors between 2% - 4% for quartz, HOPG, and sapphire. The measurements also showed reasonable agreement with literature values for the thermal conductivity tensor elements. These results demonstrate the feasibility of the extended BO-FDTR technique presented in this work, which can facilitate anisotropic thermal conductivity measurements for a broader class of materials with arbitrary in-plane anisotropy.

## Acknowledgements

The authors gratefully acknowledge funding from the Directed Energy - Joint Transition Office (DE-JTO) and the Office of Naval Research. This research used the Savio computational cluster resource provided by the Berkeley Research Computing program at the University of California, Berkeley (supported by the UC Berkeley Chancellor, Vice Chancellor for Research, and Chief Information Officer). The authors also would like to thank Josh Wilbur for helpful discussions and Jason Wu for training on electron beam evaporation.

# References

[1] B. Sun, G. Haunschild, C. Polanco, J. Ju, L. Lindsay, G. Koblmuller, Y.K. Koh, Dislocation-induced thermal transport anisotropy in single-crystal group-III nitride films, Nat. Mater., 18(2) (2019) 136-140.

[2] L.K. Li, Y.J. Yu, G.J. Ye, Q.Q. Ge, X.D. Ou, H. Wu, D.L. Feng, X.H. Chen, Y.B. Zhang, Black phosphorus field-effect transistors, Nat. Nanotechnol., 9(5) (2014) 372-377.

[3] X.A. Yan, B. Poudel, Y. Ma, W.S. Liu, G. Joshi, H. Wang, Y.C. Lan, D.Z. Wang, G. Chen, Z.F. Ren, Experimental Studies on Anisotropic Thermoelectric Properties and Structures of n-Type $Bi_2Te_{2.7}Se_{0.3}$, Nano Lett., 10(9) (2010) 3373-3378.

[4] L.D. Zhao, S.H. Lo, Y.S. Zhang, H. Sun, G.J. Tan, C. Uher, C. Wolverton, V.P. Dravid, M.G. Kanatzidis, Ultralow thermal conductivity and high thermoelectric figure of merit in SnSe crystals, Nature, 508(7496) (2014) 373-377.

[5] J.H. Chu, J.G. Analytis, K. De Greve, P.L. McMahon, Z. Islam, Y. Yamamoto, I.R. Fisher, In-Plane Resistivity Anisotropy in an Underdoped Iron Arsenide Superconductor, Science, 329(5993) (2010) 824-826.

[6] A.T. Wieg, Y. Kodera, Z. Wang, T. Imai, C. Dames, J.E. Garay, Visible photoluminescence in polycrystalline terbium doped aluminum nitride (Tb:AlN) ceramics with high thermal conductivity, Appl. Phys. Lett., 101(11) (2012) 111903.

[7] A.T. Wieg, Y. Kodera, Z. Wang, C. Dames, J.E. Garay, Thermomechanical properties of rare-earth-doped AlN for laser gain media: The role of grain boundaries and grain size, Acta Mater., 86 (2015) 148-156.

[8] T. Borca-Tasciuc, A.R. Kumar, G. Chen, Data reduction in 3 omega method for thin-film thermal conductivity determination, Rev. Sci. Instrum., 72(4) (2001) 2139-2147.

[9] T. Tong, A. Majumdar, Reexamining the 3-omega technique for thin film thermal characterization, Rev. Sci. Instrum., 77(10) (2006) 104902.

[10] A.T. Ramu, J.E. Bowers, Analysis of the "3-Omega" method for substrates and thick films of anisotropic thermal conductivity, J. Appl. Phys., 112(4) (2012) 043516.

[11] S. Lee, F. Yang, J. Suh, S.J. Yang, Y. Lee, G. Li, H.S. Choe, A. Suslu, Y.B. Chen, C. Ko, J. Park, K. Liu, J.B. Li, K. Hippalgaonkar, J.J. Urban, S. Tongay, J.Q. Wu, Anisotropic in-plane thermal conductivity of black phosphorus nanoribbons at temperatures higher than 100 K, Nat. Commun., 6(1) (2015) 1-7.

[12] S. Kwon, J.L. Zheng, M.C. Wingert, S. Cui, R.K. Chen, Unusually High and Anisotropic Thermal Conductivity in Amorphous Silicon Nanostructures, Acs Nano, 11(3) (2017) 2470-2476.

[13] J. Yang, C. Maragliano, A.J. Schmidt, Thermal property microscopy with frequency domain thermoreflectance, Rev. Sci. Instrum., 84(10) (2013) 104904.

[14] J. Yang, E. Ziade, C. Maragliano, R. Crowder, X.Y. Wang, M. Stefancich, M. Chiesa, A.K. Swan, A.J. Schmidt, Thermal conductance imaging of graphene contacts, J. Appl. Phys., 116(2) (2014) 023515.

[15] V.V. Medvedev, J. Yang, A.J. Schmidt, A.E. Yakshin, R.W.E. van de Kruijs, E. Zoethout, F. Bijkerk, Anisotropy of heat conduction in Mo/Si multilayers, J. Appl. Phys., 118(8) (2015) 085101.

[16] D. Rodin, S.K. Yee, Simultaneous measurement of in-plane and through-plane thermal conductivity using beam-offset frequency domain thermoreflectance, Rev. Sci. Instrum., 88(1) (2017) 014902.

[17] M. Rahman, M. Shahzadeh, P. Braeuninger-Weimer, S. Hofmann, O. Hellwig, S. Pisana, Measuring the thermal properties of anisotropic materials using beam-offset frequency domain thermoreflectance, J. Appl. Phys., 123(24) (2018) 245110.

[18] A.J. Schmidt, X.Y. Chen, G. Chen, Pulse accumulation, radial heat conduction, and anisotropic thermal conductivity in pump-probe transient thermoreflectance, Rev. Sci. Instrum., 79(11) (2008) 114902.

[19] J.P. Feser, D.G. Cahill, Probing anisotropic heat transport using time-domain thermoreflectance with offset laser spots, Rev. Sci. Instrum., 83(10) (2012) 104901.

[20] J.P. Feser, J. Liu, D.G. Cahill, Pump-probe measurements of the thermal conductivity tensor for materials lacking in-plane symmetry, Rev. Sci. Instrum., 85(10) (2014) 104903.

[21] P.Q. Jiang, X. Qian, R.G. Yang, A new elliptical-beam method based on time-domain thermoreflectance (TDTR) to measure the in-plane anisotropic thermal conductivity and its comparison with the beam-offset method, Rev. Sci. Instrum., 89(9) (2018) 094902.
[22] M. Li, J.S. Kang, Y.J. Hu, Anisotropic thermal conductivity measurement using a new Asymmetric-Beam Time-Domain Thermoreflectance (AB-TDTR) method, Rev. Sci. Instrum., 89(8) (2018) 084901.
[23] X. Qian, Z. Ding, J. Shin, A.J. Schmidt, G. Chen, Accurate measurement of in-plane thermal conductivity of layered materials without metal film transducer using frequency domain thermoreflectance, Rev. Sci. Instrum., 91(6) (2020) 064903.
[24] K.T. Regner, S. Majumdar, J.A. Malen, Instrumentation of broadband frequency domain thermoreflectance for measuring thermal conductivity accumulation functions, Rev. Sci. Instrum., 84(6) (2013) 064901.
[25] D.G. Cahill, Analysis of heat flow in layered structures for time-domain thermoreflectance, Rev. Sci. Instrum., 75(12) (2004) 5119-5122.
[26] C. Cardenas, D. Fabris, S. Tokairin, F. Madriz, C.Y. Yang, Thermoreflectance Measurement of Temperature and Thermal Resistance of Thin Film Gold, J. Heat Trans.-T Asme, 134(11) (2012).
[27] D.W. Hahn, M.N. Özisik, Heat conduction, John Wiley & Sons, 2012.
[28] L. Onsager, Reciprocal relations in irreversible processes. I., Phys. Rev., 37(4) (1931) 405-426.
[29] J.A. Malen, K. Baheti, T. Tong, Y. Zhao, J.A. Hudgings, A. Majumdar, Optical Measurement of Thermal Conductivity Using Fiber Aligned Frequency Domain Thermoreflectance, J. Heat Trans.-T Asme, 133(8) (2011).
[30] Y. Touloukian, E. Buyco, Thermophysical Properties of Matter, Vol. 4, Specific Heat, IFI/Plenum, New York, (1970).
[31] A.J. Schmidt, R. Cheaito, M. Chiesa, A frequency-domain thermoreflectance method for the characterization of thermal properties, Rev. Sci. Instrum., 80(9) (2009) 094901.
[32] W. Desorbo, W.W. Tyler, The Specific Heat of Graphite from 13-Degrees-K to 300-Degrees-K, J. Chem. Phys., 21(10) (1953) 1660-1663.
[33] J.H. Lienhard IV, J.H. Lienhard, V: A Heat Transfer Textbook. Version 1.23, in, Phlogiston Press, Lexington, MA, 2005.
[34] R.E. Newnham, Properties of materials: anisotropy, symmetry, structure, Oxford University Press on Demand, 2005.
[35] E.R. Dobrovinskaya, L.A. Lytvynov, V. Pishchik, Application of sapphire, in: Sapphire, Springer, 2009, pp. 55-153.
[36] D.G. Cahill, S.-M. Lee, T.I. Selinder, Thermal conductivity of κ-$Al_2O_3$ and α-$Al_2O_3$ wear-resistant coatings, J. Appl. Phys., 83(11) (1998) 5783-5786.
[37] C. Monachon, L. Weber, C. Dames, Thermal Boundary Conductance: A Materials Science Perspective, Annu. Rev. Mater. Res., 46 (2016) 433-463.
[38] P.Q. Jiang, X. Qian, R.G. Yang, Time-domain thermoreflectance (TDTR) measurements of anisotropic thermal conductivity using a variable spot size approach, Rev. Sci. Instrum., 88(7) (2017) 074901.
[39] J.N. Reddy, D.K. Gartling, The finite element method in heat transfer and fluid dynamics, CRC press, 2010.